\documentclass[11pt]{article}

\usepackage[margin=1in]{geometry}
\usepackage{microtype}
\usepackage{booktabs}
\usepackage{graphicx}
\usepackage{amsmath}
\usepackage{natbib}
\usepackage{hyperref}
\usepackage{listings}
\usepackage{xcolor}

\hypersetup{colorlinks=true, linkcolor=blue!60!black,
            citecolor=blue!60!black, urlcolor=blue!60!black}

\newcommand{\sig}{$^{\ast}$}

\graphicspath{{./}{../figs/}}

\title{\texttt{lifelines-hc}: Higher Criticism testing for sparse
  non-proportional hazard departures in Python}

\author{%
  Alon Kipnis\thanks{School of Computer Science, Reichman University,
    Herzliya 4610101, Israel. Corresponding author:
    \href{mailto:alon.kipnis@runi.ac.il}{alon.kipnis@runi.ac.il}}
  \and
  Ben Galili\thanks{Department of Computer Science, Technion -- Israel
    Institute of Technology, Haifa 3200003, Israel}
  \and
  Zohar Yakhini\thanks{School of Computer Science, Reichman University,
    Herzliya 4610101, Israel; and Department of Computer Science,
    Technion -- Israel Institute of Technology, Haifa 3200003, Israel}%
}

\date{}

\begin{document}
\maketitle

\begin{abstract}
\textbf{Motivation:}
The log-rank test is a standard tool for two-sample survival comparison and has good power against the widely studied proportional-hazards alternatives. However, \citet{kipnis_biometrika} showed that it can lose power against sparse hazard departures, where the hazard difference is concentrated in a small number of time intervals whose locations along the follow-up are unknown a priori. Weighted log-rank tests, such as Gehan-Wilcoxon, Tarone-Ware, Peto-Prentice, and Fleming-Harrington, each impose a pre-specified temporal emphasis. Combination procedures such as MaxCombo \citep{lin2020maxcombo} and the Yang--Prentice short-term/long-term hazard-ratio model \citep{yang2005,yang2010improved} relax that choice, but still scan only a small dictionary of global temporal shapes, and are therefore also insensitive to sparse departures occurring outside those shapes.

\textbf{Results:}
We introduce \texttt{lifelines-hc}, a Python package extending the
\texttt{lifelines} survival library with the HCHG test: Higher Criticism
(HC) applied to per-interval HyperGeometric (HG) p-values from
\citet{kipnis_biometrika}. 
HCHG is a test for right-censored two-sample survival data. It detects sparse hazard departures at unknown locations without committing to any temporal pattern. Across four biological case studies
— an immuno-oncology trial (CheckMate~057 PFS, $n=582$), a targeted
kinase-inhibitor trial in metastatic prostate cancer (COMET-1 OS,
$n=1028$), an adjuvant bisphosphonate trial (AZURE DFS, $n=3359$),
and the
Copenhagen Study Group for Liver Diseases (CSL) liver-cirrhosis trial with fully public individual patient data ($n=446$) — HCHG achieves $p \leq 0.014$ in every dataset, while the log-rank test is non-significant throughout ($p \geq 0.26$). MaxCombo and the Yang--Prentice adaptive log-rank test detect the delayed-benefit crossing on CheckMate~057 ($p < 0.001$) but are non-significant on the remaining three datasets ($p \geq 0.24$), where the departure is sparse or multi-window rather than a smooth short-term/long-term hazard-ratio pattern.\\

\textbf{Availability and Implementation:}
\texttt{lifelines-hc} is freely available under the MIT license at
\url{https://github.com/alonkipnis/lifelines-hc} and via
\texttt{pip install lifelines-hc}; it requires Python~${\geq}\,3.8$ and
\texttt{lifelines}~${\geq}\,0.29$. Full analysis scripts and per-method
results tables are provided as Supplementary Data.%
\end{abstract}


\vspace{1em}
\section{Introduction}

The Kaplan-Meier estimator \citep{kaplan1958} and the log-rank test
\citep{mantel1966} are the workhorses of two-sample survival analysis in
biomedical research. The log-rank statistic has good power against a wide
range of useful alternatives, particularly proportional-hazards (PH)
alternatives \citep{schoenfeld1981}. Nevertheless,
\citet{kipnis_biometrika} showed that the log-rank test is not optimal
against \emph{sparse} hazard departures, in which the hazard difference is
concentrated in a small, unknown subset of time intervals.

Such localized or non-proportional departures arise naturally in several
biomedical settings. Immune checkpoint inhibitors may require an
immune-priming phase before producing durable tumour suppression, leading
to delayed treatment effects and crossing survival curves; for example,
in the second-line non-squamous NSCLC trial CheckMate~057, nivolumab
improved overall survival relative to docetaxel, while the
progression-free survival curves crossed, consistent with delayed benefit
\citep{borghaei2015}. Targeted therapies can also produce endpoint-specific
benefits over a limited follow-up window that are diluted in global
survival summaries: in the COMET-1 mCRPC trial, cabozantinib markedly
improved bone scan response at week~12 and radiographic
progression-free survival, but did not significantly improve overall
survival \citep{smith2016}. Subgroup-dependent biology can similarly
generate composite, time-varying hazard differences: in the AZURE trial
of adjuvant zoledronic acid, no overall benefit was observed in the full
population, but benefit was seen among women with established menopause,
consistent with an effect mediated by the postmenopausal bone
microenvironment \citep{coleman2011,coleman2014}.
Time-varying treatment effects can likewise produce multi-window hazard
departures: in the CSL liver-cirrhosis trial, prednisone was harmful early
(infection and fluid retention) but beneficial later through
anti-inflammatory action, yielding survival curves that cross more than once
\citep{csl_data}. This canonical non-proportional-hazards dataset is widely
used to illustrate time-varying-coefficient models and provides fully public
individual patient data via the CRAN \texttt{timereg} package
\citep{csl_data} and the SurvSet repository \citep{survset2022}.

The need for log-rank alternatives in survival analysis is well recognized,
and several weighted log-rank tests have been developed for non-proportional-hazards
settings \citep{gehan1965,peto1972,taroneware1977,harrington1982,flemingharrington1991}.
Each applies a temporal weight function, emphasizing early, intermediate,
or late events. Each of these tests achieves good power when the signal matches the intended temporal pattern. However, choosing an appropriate weight requires knowing in advance \emph{where} the hazard departure will occur. When the departure is sparse but its location is unknown a priori, all weighted log-rank tests can miss it simultaneously, as we demonstrate below.

Contemporary practice therefore combines several Fleming--Harrington weights rather than committing to one. The MaxCombo test \citep{lin2020maxcombo,ristl2021nph} takes the maximum of $\mathrm{FH}(\rho,\gamma)$ statistics over the dictionary $(\rho,\gamma)\in\{(0,0),(0,1),(1,0),(1,1)\}$ --- log-rank, late, early, and middle emphasis --- and obtains a multiplicity-adjusted $p$-value from the estimated multivariate normal joint distribution of the component $Z$-statistics. The Yang--Prentice short-term/long-term hazard-ratio model \citep{yang2005} interpolates between an immediate hazard ratio $\theta_1$ and an asymptotic ratio $\theta_2$, covering delayed, diminishing, and crossing hazards with two parameters; the associated adaptive weighted log-rank test \citep{yang2010improved} is a leading model-based competitor. Both procedures are powerful against the global early-versus-late patterns they were designed for, particularly delayed benefit in immuno-oncology. They remain, however, averaging methods over a small dictionary of global temporal shapes, and can miss sparse or multi-window departures whose locations are unknown a priori.

\citet{kipnis_biometrika} proposed the HCHG test — the Higher Criticism
(HC) statistic of \citet{donoho2004} applied to per-interval
HyperGeometric (HG) p-values — specifically for this setting. HCHG
divides the follow-up into equal-width intervals, derives a hypergeometric
p-value for the event-count comparison in each interval under the null of
equal group-specific hazards, and applies the HC statistic to test whether
the body of these p-values is collectively implausible
under the global null. Because HCHG does not pre-specify any temporal
weight, it retains power for sparse departures wherever they occur along
the follow-up axis.

We present \texttt{lifelines-hc}, a Python package that implements HCHG and integrates with the popular \texttt{lifelines} library \citep{davidsonpilon2019}.

\section{The \texttt{lifelines-hc} package}

\subsection*{The HCHG statistic}

HCHG \citep{kipnis_biometrika} combines two components: hypergeometric
interval p-values and the HC statistic of \citet{donoho2004,donoho2015special}.

\textbf{Hypergeometric interval p-values.}
Given two groups with right-censored observations $(T_i, E_i, G_i)$,
partition the follow-up range into $K$ equal-width intervals. For
interval $k$, let $n_k$ be the total events, $r_k^A$ and $r_k^B$ the
subjects at risk in groups A and B, and $x_k$ the events in group B.
Under the null of equal group-specific hazards,
\begin{equation}\label{eq:hypergeom}
  x_k \;\Big|\; n_k,\, r_k^A,\, r_k^B
  \;\sim\;
  \mathrm{Hypergeometric}\!\left(r_k^A + r_k^B,\; r_k^B,\; n_k\right).
\end{equation}
This yields a one-sided p-value $p_k$ for each interval using an exact test.

\textbf{HC aggregation.}
Ordering the $K$ p-values as $p_{(1)} \leq \cdots \leq p_{(K)}$, the
HC statistic \citep{donoho2004,donoho2015special} is applied to this collection:
\begin{equation}\label{eq:hc}
  \mathrm{HCHG}_K
  \;=\;
  \max_{1 \leq j \leq \lfloor\gamma K\rfloor}
  \frac{j/K - p_{(j)}}{\sqrt{p_{(j)}\!\left(1 - p_{(j)}\right)/K}},
\end{equation}
where $\gamma \in (0,1)$ (default $0.2$) restricts scanning to the most
extreme fraction of interval p-values \citep{donoho2004,donoho2015special}. A global
p-value is obtained by permuting group labels
($n_\mathrm{perms} = 2000$ in the analyses reported here). For the two-sided alternative,
the minimum of the two one-sided interval p-values replaces $p_k$
in~(\ref{eq:hc}).

\subsection*{Software design and API}

\texttt{lifelines-hc} exposes a function-based API consistent with
\texttt{lifelines.statistics}, accepting NumPy arrays of event times and
indicators:

\begin{lstlisting}
from lifelines_hc import (higher_criticism_test, suspected_deviations)

result = higher_criticism_test(
    T_A, T_B,
    event_observed_A=E_A, event_observed_B=E_B,
    n_intervals_to_pool=100, n_permutations=2000,
    alternative='both', gamma=0.2, seed=42)

print(result.p_value)       # permutation-calibrated p-value
print(result.test_statistic)  # HCHG statistic value
\end{lstlisting}

\noindent
The companion \texttt{suspected\_deviations()} function returns a
\texttt{pandas.DataFrame} of per-interval p-values with a
\texttt{suspected} boolean column marking HCHG-flagged intervals,
enabling overlay of diagnostic shading on Kaplan-Meier plots. 
The package additionally ships with plotting utilities
(\texttt{plot\_km\_with\_hc}, \texttt{plot\_pvalue\_profile}) and a
convenience wrapper \texttt{run\_all\_tests()} that benchmarks HCHG, log-rank, the four weighted alternatives, MaxCombo, and Yang--Prentice in a single
call, returning a \texttt{pandas.DataFrame} for easy comparison.

\begin{figure}[t!]
  \centering
  \includegraphics[width=\textwidth]{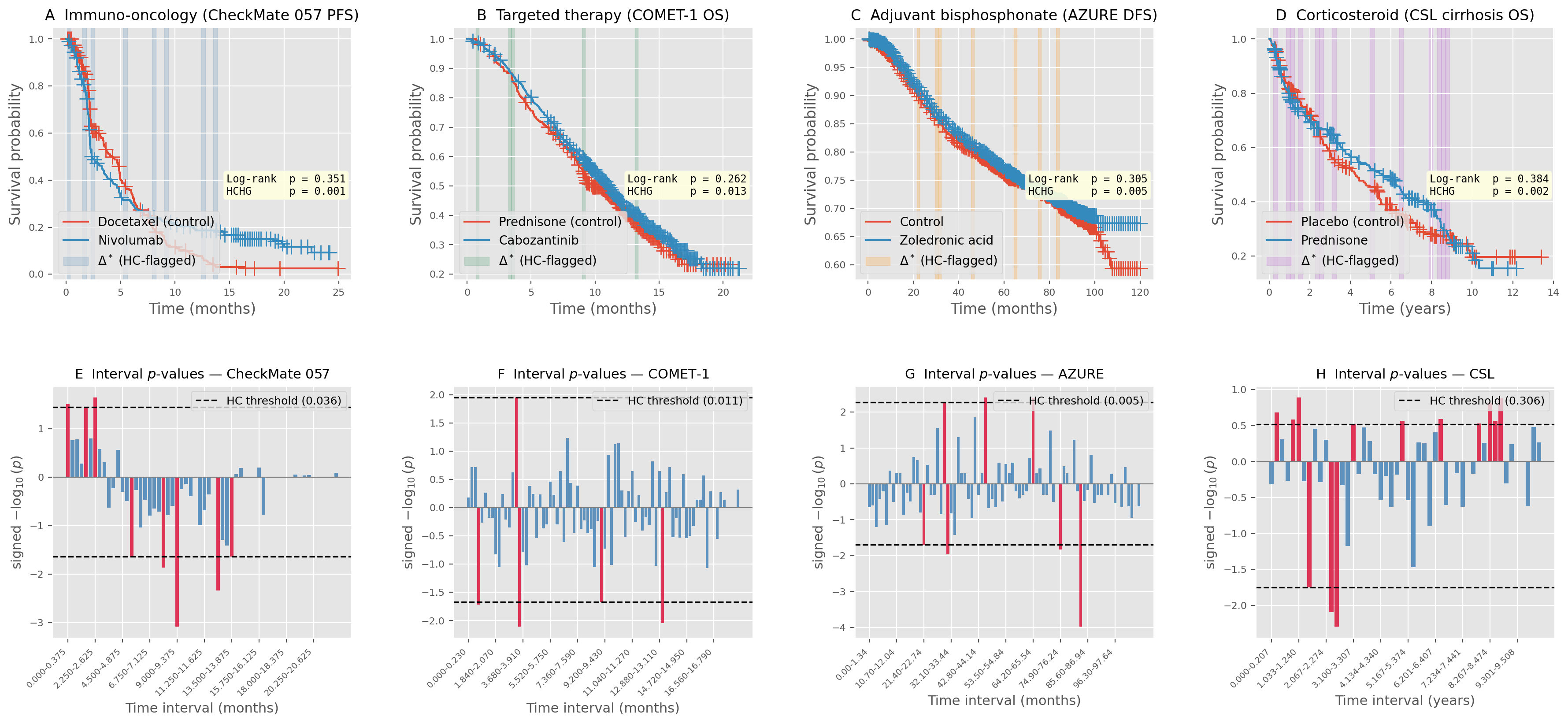}
  \caption{%
    \textbf{Higher Criticism detects non-proportional hazard departures
    across four clinical datasets with distinct temporal patterns.}
    Top row (A--D): Kaplan-Meier curves with HCHG-flagged intervals
    (shaded). Bottom row (E--H): per-interval \emph{signed}
    $-\!\log_{10}$(hypergeometric p-value).     Upward bars mark excess
    events in the treatment arm, downward bars mark excess in the control arm.
    Dashed lines are the per-direction HCHG thresholds; bars reaching
    beyond either line are highlighted in red and coincide exactly with
    the intervals shaded above.
    \textbf{(A,E)}~CheckMate~057 PFS (nivolumab vs.\ docetaxel, $n=582$);
    log-rank $p=0.351$, HCHG $p=0.001$. Curves cross at $\sim\!3$ months
    before nivolumab gains a durable late advantage.
    \textbf{(B,F)}~COMET-1 OS (cabozantinib vs.\ prednisone, $n=1028$);
    log-rank $p=0.262$, HCHG $p=0.014$. Early OS advantage (months~3--7)
    during bone-lesion response, fading as resistance develops.
    \textbf{(C,G)}~AZURE DFS (zoledronic acid vs.\ control, $n=3359$);
    log-rank $p=0.305$, HCHG $p=0.005$. Menopause-dependent benefit
    accumulates in months~20--60.
    \textbf{(D,H)}~CSL OS (prednisone vs.\ placebo, liver cirrhosis,
    $n=446$); log-rank $p=0.384$, HCHG $p=0.002$. Time-varying
    corticosteroid effect with multiple curve crossings; \emph{fully
    public} raw individual patient data (CRAN \texttt{timereg} / SurvSet).
    In every dataset the log-rank test is non-significant while HCHG is
    significant. MaxCombo and Yang--Prentice are significant only on the
    delayed-benefit crossing in (A,E); see Supplementary Data.
    For the first three, individual patient data were
    reconstructed via \citet{guyot2012}.%
  }
  \label{fig:main}
\end{figure}

\section{Results}

We demonstrate \texttt{lifelines-hc} on four biological datasets with
distinct non-proportional-hazards (PH) structures (Figure~\ref{fig:main}), benchmarking HCHG
against the log-rank test, four weighted alternatives, the MaxCombo combination test, and the Yang--Prentice adaptive log-rank test.
Individual patient data for the first three trials were reconstructed from
published Kaplan-Meier curves; the CSL trial provides publicly available individual patient data.
Full per-method $p$-value tables are provided in Supplementary Data.

\subsection*{Immuno-oncology: crossing survival curves}

CheckMate~057 \citep{borghaei2015} randomised 582 patients with advanced
non-squamous NSCLC to nivolumab ($n=292$) or docetaxel ($n=290$).
Progression-free survival showed crossing curves: docetaxel provided a
short-term advantage during immune priming, subsequently overtaken by the
durable nivolumab response (Figure~\ref{fig:main}A). Individual patient
data were reconstructed from the published Kaplan-Meier curve via the
algorithm of \citet{guyot2012} as implemented in the \texttt{kmdata}
R package.

The log-rank test is non-significant ($p = 0.351$) as the early excess
and late benefit partially cancel. Among the weighted alternatives,
Gehan-Wilcoxon ($p = 0.041$) and Peto-Prentice ($p = 0.049$) reach
borderline significance by detecting the early crossing, but
Tarone-Ware ($p = 0.358$) and Fleming-Harrington\,(1,1) ($p = 0.256$)
do not. This is the delayed-benefit pattern MaxCombo and Yang--Prentice
are designed for: MaxCombo is highly significant ($p = 2.5 \times 10^{-5}$),
driven by the late-weighted $\mathrm{FH}(0,1)$ member ($Z = 4.27$), and
the Yang--Prentice adaptive log-rank test is likewise significant
($p = 0.00055$), with estimated short-term hazard ratio $\hat\theta_1 = 2.33$
(nivolumab excess during priming) and long-term ratio $\hat\theta_2 = 0.44$
(durable nivolumab benefit). HCHG achieves $p = 0.001$ and characterises
the full crossing structure by flagging both the early divergence and late
separation windows (Figure~\ref{fig:main}A,E).

\subsection*{Targeted therapy in prostate cancer: a transient early benefit}

The COMET-1 trial \citep{smith2016} randomised 1028 patients with
chemotherapy-pretreated mCRPC to cabozantinib ($n=682$) or prednisone
($n=346$) in a 2:1 ratio. Overall survival showed no significant
log-rank difference ($p = 0.262$). Cabozantinib, by inhibiting
MET and VEGFR2 kinases, significantly improved bone scan response at
week~12 (a pre-specified secondary endpoint), confirming genuine
biological activity against bone-metastatic disease.
This short-term activity did not, however, translate into a sustained overall survival (OS) benefit. 
The result is a temporally concentrated early OS advantage
(months~3--7) that is subsequently lost (Figure~\ref{fig:main}B,F).

All four weighted alternatives are non-significant ($p \geq 0.19$):
the early-weighted Gehan-Wilcoxon ($p = 0.193$) and Peto-Prentice
($p = 0.206$) come closest but are diluted by the null period after
the bone-response window. MaxCombo ($p = 0.338$) and Yang--Prentice
($p = 0.262$) likewise miss the signal: a transient few-month advantage
is not a global early-versus-late hazard-ratio shape. HCHG ($p = 0.014$) aggregates evidence from the specific early intervals
where the transient effect is concentrated.

\subsection*{Adjuvant bisphosphonate therapy: a menopause-dependent effect}

The AZURE trial \citep{coleman2011,coleman2014} randomised 3359 early-stage
breast cancer patients to standard adjuvant therapy with ($n=1681$) or
without ($n=1678$) zoledronic acid. Disease-free survival showed no
significant log-rank difference ($p = 0.305$) because the drug's
bone-microenvironment benefit depends on the low-oestrogen state of
postmenopause, which was absent in the many premenopausal patients
enrolled at baseline. As these patients progressively transitioned to
postmenopause during the decade-long follow-up, the hazard difference
accumulated in a specific mid-to-late window (months 20--60) rather than
uniformly (Figure~\ref{fig:main}C,G).

All four weighted alternatives are non-significant ($p \geq 0.28$):
no single temporal weight captures a pattern distributed across this
window while remaining unaffected by the null early period. MaxCombo
($p = 0.401$) and Yang--Prentice ($p = 0.308$) are similarly
non-significant. HCHG
($p = 0.005$) detects this scattered but collectively significant signal.
The result is validated by published subgroup analyses showing significant disease-free survival (DFS) benefit in postmenopausal patients \citep{coleman2011,coleman2014},
confirming that HCHG is detecting a genuine biological effect.

\subsection*{Corticosteroid therapy in liver cirrhosis: a time-varying effect}

The CSL trial \citep{csl_data} randomised 446 patients with histologically
verified liver cirrhosis to prednisone ($n=226$) or placebo ($n=220$) in
Copenhagen (1962--69), with 270 deaths over follow-up of up to
$\sim$13 years. Unlike the preceding case studies, the raw individual
patient data are genuinely public---distributed in the CRAN \texttt{timereg}
package (dataset \texttt{csl}) and the open SurvSet repository
\citep{survset2022}---rather than reconstructed from a published
Kaplan-Meier curve.

Corticosteroids exert a time-varying effect in cirrhosis: early harm through
immunosuppression, infection, and fluid retention is followed by later
anti-inflammatory benefit in actively inflamed subgroups. The survival
curves cross more than once, with the treatment effect concentrated in a
few non-contiguous windows (Figure~\ref{fig:main}D,H).

The log-rank test is non-significant ($p = 0.384$), and all four weighted
alternatives are as well ($p \geq 0.13$): Gehan-Wilcoxon ($p = 0.51$),
Tarone-Ware ($p = 0.38$), Peto-Prentice ($p = 0.45$), and
Fleming-Harrington\,(1,1) ($p = 0.14$). MaxCombo ($p = 0.236$) and
Yang--Prentice ($p = 0.393$) also fail: no single temporal weight or
two-parameter short-term/long-term hazard ratio captures a multi-window,
sign-changing departure. HCHG achieves
$p = 0.002$, demonstrating detection of sparse non-proportional-hazards departures on genuinely public raw data.

To summarise, across the four case studies, HCHG achieves $p \leq 0.014$ in every domain,
while the log-rank test is non-significant throughout ($p \geq 0.26$). MaxCombo and Yang--Prentice are significant only on CheckMate~057, the delayed-benefit crossing they were designed to detect; on COMET-1, AZURE, and CSL they return $p \geq 0.24$. The
two borderline significances for Gehan-Wilcoxon ($p = 0.041$) and
Peto-Prentice ($p = 0.049$) in CheckMate~057 are the only other instances where
any weighted alternative reaches $p < 0.05$. HCHG is therefore complementary to MaxCombo and Yang--Prentice: it matches them on global early-versus-late non-proportionality and retains power for sparse or multi-window departures that a small Fleming--Harrington dictionary cannot represent.

\section{Conclusion}

\texttt{lifelines-hc} provides an accessible Python implementation of
HCHG for two-sample right-censored survival data. HCHG is specifically
designed for sparse hazard departures at unknown temporal locations:
unlike weighted log-rank alternatives, MaxCombo, and the Yang--Prentice
adaptive test, it requires no a priori
specification of where the hazard difference will occur and retains
power for any concentrated departure along the follow-up axis. The
package integrates transparently with \texttt{lifelines} workflows and
returns diagnostic interval-level information identifying which time
windows drive the signal.



\section*{Conflict of Interest}
None declared.

\section*{Funding}
The work of Alon Kipnis was supported in part by the US-Israel Binational Science Foundation (BSF) grant 2022124.

\section*{Data availability}
Individual patient data for CheckMate~057, COMET-1, and AZURE were
reconstructed from published Kaplan-Meier curves using the algorithm of
\citet{guyot2012} as implemented in the \texttt{kmdata} R package
(\url{https://github.com/raredd/kmdata}). CSL liver-cirrhosis individual
patient data are fully public via the CRAN \texttt{timereg} package
(dataset \texttt{csl}) and the SurvSet repository \citep{survset2022}
(\url{https://github.com/ErikinBC/SurvSet}).
All analysis code is available at
\url{https://github.com/alonkipnis/lifelines-hc/tree/v0.1.2/AppNote}.


\appendix
\section*{Supplementary material}
\addcontentsline{toc}{section}{Supplementary material}
\renewcommand{\thesection}{S\arabic{section}}
\setcounter{section}{0}
\renewcommand{\thetable}{S\arabic{table}}
\setcounter{table}{0}

\noindent
This appendix provides (i)~the full analysis workflow and scripts used to
reproduce the case studies, and (ii)~complete per-method test statistics
and $p$-values for all four datasets analysed in the main text.

\section{Analysis scripts and data}

All scripts live in the \texttt{AppNote/} directory of the
\texttt{lifelines-hc} repository
(\url{https://github.com/alonkipnis/lifelines-hc/tree/v0.1.2/AppNote}).
They require Python~${\geq}\,3.8$, \texttt{lifelines-hc}, and
\texttt{lifelines}~${\geq}\,0.29$.
The MaxCombo and Yang--Prentice comparisons additionally require
\texttt{R} with the CRAN packages \texttt{nph} \citep{ristl2021nph}
and \texttt{YPmodel} \citep{yang2005,yang2010improved}.

\subsection*{Workflow}

\begin{enumerate}
  \item \textbf{Download trial IPD.}
        \texttt{01\_get\_kmdata.py} exports reconstructed individual patient
        data (IPD) from the \texttt{kmdata} R package
        \citep{guyot2012} for CheckMate~057 \citep{borghaei2015},
        COMET-1 \citep{smith2016}, and AZURE
        \citep{coleman2011,coleman2014}.
        CSL liver-cirrhosis IPD are downloaded directly from the CRAN
        \texttt{timereg} package and the SurvSet repository
        \citep{csl_data,survset2022} (see \texttt{run\_csl.py}).
  \item \textbf{Run per-dataset analyses.}
        Each \texttt{run\_*.py} script loads one dataset, calls
        \texttt{run\_all\_tests()} in \texttt{utils.py}, writes a results
        CSV to \texttt{results/}, and saves diagnostic figures to
        \texttt{figs/}.  \texttt{run\_all\_tests()} also invokes
        \texttt{competitors.R} for MaxCombo and Yang--Prentice.
  \item \textbf{Build the composite figure.}
        \texttt{make\_figure.py} assembles Figure~1 of the main text from
        the four domain scripts.
\end{enumerate}

\subsection*{Script index}

\begin{table}[h]
  \centering
  \caption{Analysis scripts for the four case studies.}
  \label{tab:scripts}
  \small
  \begin{tabular}{@{}llp{6.8cm}@{}}
    \toprule
    Script & Dataset & Role \\
    \midrule
    \texttt{01\_get\_kmdata.py} & CheckMate~057, COMET-1, AZURE &
      Download IPD via \texttt{kmdata} (Guyot algorithm). \\
    \texttt{run\_immuno\_oncology.py} & CheckMate~057 PFS ($n=582$) &
      Nivolumab vs.\ docetaxel; crossing curves. \\
    \texttt{run\_comet.py} & COMET-1 OS ($n=1028$) &
      Cabozantinib vs.\ prednisone; transient early benefit. \\
    \texttt{run\_azure.py} & AZURE DFS ($n=3359$) &
      Zoledronic acid vs.\ control; menopause-dependent effect. \\
    \texttt{run\_csl.py} & CSL OS ($n=446$) &
      Prednisone vs.\ placebo; public raw IPD, multi-window effect. \\
    \texttt{utils.py} & --- &
      Shared \texttt{run\_all\_tests()}, plotting, and table utilities. \\
    \texttt{competitors.R} & --- &
      MaxCombo (\texttt{nph::logrank.maxtest}) and Yang--Prentice
      adaptive log-rank (\texttt{YPmodel::YPmodel.adlgrk}). \\
    \texttt{make\_figure.py} & All four &
      Composite Figure~1 panels (A--H). \\
    \bottomrule
  \end{tabular}
\end{table}

\subsection*{Reproduction commands}

\begin{lstlisting}
cd AppNote
python 01_get_kmdata.py --trials Checkmate057_1C COMET1_2A AZURE_2A
python run_immuno_oncology.py
python run_comet.py
python run_azure.py
python run_csl.py          # requires SurvSet or AppNote/data/CSL.csv
python make_figure.py
# MaxCombo / Yang--Prentice: requires R packages nph and YPmodel
#   install.packages(c("nph", "YPmodel"))
# invoked automatically from run_all_tests() via competitors.R
\end{lstlisting}

Each \texttt{run\_*.py} script writes
\texttt{results/<domain>\_test\_results.csv} containing one row per
competing method (statistic and permutation- or asymptotic-based
$p$-value).

\section{Data sources}

Table~\ref{tab:data} lists the original clinical studies, endpoints, and
how each dataset was obtained.  For CheckMate~057, COMET-1, and AZURE,
only published Kaplan--Meier curves were available; IPD were
reverse-engineered with the algorithm of \citet{guyot2012} and distributed
in the \texttt{kmdata} package (\url{https://github.com/raredd/kmdata}).
The CSL trial is the only case study with genuinely public raw IPD.

\begin{table}[h]
  \centering
  \caption{Original studies and data acquisition for the four case studies.}
  \label{tab:data}
  \small
  \begin{tabular}{@{}p{2.1cm}p{1.2cm}rp{4.6cm}@{}}
    \toprule
    Trial & Endpoint & $n$ & Source and acquisition \\
    \midrule
    CheckMate~057 &
      PFS & 582 &
      \citet{borghaei2015}; IPD reconstructed from published
      Kaplan--Meier curves via \citet{guyot2012} (\texttt{kmdata}). \\
    COMET-1 &
      OS & 1028 &
      \citet{smith2016}; IPD reconstructed from published
      Kaplan--Meier curves via \citet{guyot2012} (\texttt{kmdata}). \\
    AZURE &
      DFS & 3359 &
      \citet{coleman2011,coleman2014}; IPD reconstructed from published
      Kaplan--Meier curves via \citet{guyot2012} (\texttt{kmdata}). \\
    CSL &
      OS & 446 &
      Copenhagen liver-cirrhosis trial (1962--1969); raw IPD from the CRAN
      \texttt{timereg} package \citep{csl_data} and SurvSet
      \citep{survset2022}. \\
    \bottomrule
  \end{tabular}
\end{table}

\noindent
Reconstructed IPD files are written to \texttt{AppNote/data/} as
\texttt{Checkmate057\_1C.csv}, \texttt{COMET1\_2A.csv}, and
\texttt{AZURE\_2A.csv} by \texttt{01\_get\_kmdata.py}.
CSL data are written to \texttt{AppNote/data/CSL.csv} by
\texttt{run\_csl.py}.

\section{Shared analysis settings}

Unless noted otherwise, HCHG uses:
\begin{itemize}
  \item equal-width time pooling with \texttt{n\_intervals\_to\_pool}
        as listed per dataset below;
  \item two-sided interval hypergeometric $p$-values
        (\texttt{alternative='both'});
  \item HC fraction $\gamma = 0.2$;
  \item label-permutation $p$-values with the random seed fixed at 42.
\end{itemize}

The log-rank test and four weighted log-rank variants (Gehan-Wilcoxon,
Tarone-Ware, Peto-Prentice, Fleming-Harrington with $p=q=1$) use the
standard \texttt{lifelines.statistics.logrank\_test} implementation with
asymptotic $p$-values.

MaxCombo is the two-sided maximum of the Fleming--Harrington weighted
log-rank $Z$-statistics with
$(\rho,\gamma)\in\{(0,0),(0,1),(1,0),(1,1)\}$ --- the four-weight
immuno-oncology dictionary of \citet{lin2020maxcombo} --- with
$p$-value from the estimated multivariate normal joint distribution as
implemented in \texttt{nph::logrank.maxtest} \citep{ristl2021nph}
(\texttt{alternative='two.sided'}, random seed 42).  The reported
statistic is $z_{\max}=\max_j |Z_j|$.  The Yang--Prentice adaptive
weighted log-rank test of \citet{yang2010improved}, based on the
short-term/long-term hazard-ratio model of \citet{yang2005}, is computed
via \texttt{YPmodel::YPmodel.adlgrk}; the reported statistic is the
maximum of the two adaptive $|Z|$ components used internally by that
procedure.

\section{Per-method results tables}

Tables~\ref{tab:checkmate}--\ref{tab:csl} report the full comparison
for each dataset.  Asterisks mark $p < 0.05$.
Permutation-based $p$-values for HCHG are discrete
(multiples of $1/(n_\mathrm{perms}+1)$ with $n_\mathrm{perms}=2000$) and
are rounded to three decimals; ``$<0.001$'' denotes the smallest
attainable value, $1/2001$.

\subsection*{CheckMate~057 --- progression-free survival ($n=582$)}

Original trial: \citet{borghaei2015}.
IPD reconstructed from published Kaplan--Meier curves
\citep{guyot2012}.
Control: docetaxel ($n=290$); treatment: nivolumab ($n=292$).
HC settings: $K=60$ intervals, $n_\mathrm{perms}=2000$.

\begin{table}[h]
  \centering
  \caption{CheckMate~057 PFS: all competing methods.}
  \label{tab:checkmate}
  \begin{tabular}{@{}lrr@{}}
    \toprule
    Method & Statistic & $p$-value \\
    \midrule
    Log-rank                         &  0.868 & 0.351 \\
    Gehan-Wilcoxon                   &  4.169 & 0.041\sig \\
    Tarone-Ware                      &  0.844 & 0.358 \\
    Peto-Prentice                    &  3.869 & 0.049\sig \\
    Fleming-Harrington (1,1)         &  1.292 & 0.256 \\
    MaxCombo                         &  4.268 & $2.5\times 10^{-5}$\sig \\
    Yang--Prentice                   &  3.626 & 0.00055\sig \\
    HCHG                             &  1.695 & 0.001\sig \\
    \bottomrule
  \end{tabular}
\end{table}

\noindent
MaxCombo is driven by the late-weighted $\mathrm{FH}(0,1)$ member
($Z=4.27$, $p=2.0\times 10^{-5}$); $\mathrm{FH}(0,0)$, $\mathrm{FH}(1,0)$,
and $\mathrm{FH}(1,1)$ have two-sided $p=0.351$, $0.058$, and $0.256$.
The Yang--Prentice fit yields short-term hazard ratio
$\hat\theta_1=2.33$ and long-term ratio $\hat\theta_2=0.44$.

\subsection*{COMET-1 --- overall survival ($n=1028$)}

Original trial: \citet{smith2016}.
IPD reconstructed from published Kaplan--Meier curves
\citep{guyot2012}.
Control: prednisone ($n=346$); treatment: cabozantinib ($n=682$).
HC settings: $K=80$ intervals, $n_\mathrm{perms}=2000$.

\begin{table}[h]
  \centering
  \caption{COMET-1 OS: all competing methods.}
  \label{tab:comet}
  \begin{tabular}{@{}lrr@{}}
    \toprule
    Method & Statistic & $p$-value \\
    \midrule
    Log-rank                         &  1.259 & 0.262 \\
    Gehan-Wilcoxon                   &  1.693 & 0.193 \\
    Tarone-Ware                      &  1.570 & 0.210 \\
    Peto-Prentice                    &  1.602 & 0.206 \\
    Fleming-Harrington (1,1)         &  0.648 & 0.421 \\
    MaxCombo                         &  1.258 & 0.338 \\
    Yang--Prentice                   &  1.176 & 0.262 \\
    HCHG                             &  1.179 & 0.014\sig \\
    \bottomrule
  \end{tabular}
\end{table}

\subsection*{AZURE --- disease-free survival ($n=3359$)}

Original trial: \citet{coleman2011,coleman2014}.
IPD reconstructed from published Kaplan--Meier curves
\citep{guyot2012}.
Control: standard adjuvant therapy ($n=1678$); treatment: zoledronic acid
($n=1681$).  HC settings: $K=80$ intervals, $n_\mathrm{perms}=2000$.

\begin{table}[h]
  \centering
  \caption{AZURE DFS: all competing methods.}
  \label{tab:azure}
  \begin{tabular}{@{}lrr@{}}
    \toprule
    Method & Statistic & $p$-value \\
    \midrule
    Log-rank                         &  1.054 & 0.305 \\
    Gehan-Wilcoxon                   &  0.652 & 0.419 \\
    Tarone-Ware                      &  0.691 & 0.406 \\
    Peto-Prentice                    &  1.145 & 0.284 \\
    Fleming-Harrington (1,1)         &  0.277 & 0.598 \\
    MaxCombo                         &  1.072 & 0.401 \\
    Yang--Prentice                   &  1.044 & 0.308 \\
    HCHG                             &  1.513 & 0.005\sig \\
    \bottomrule
  \end{tabular}
\end{table}

\subsection*{CSL --- overall survival ($n=446$)}

Original trial: Copenhagen Study Group for Liver Diseases (1962--1969);
public IPD distributed via \citet{csl_data} and \citet{survset2022}.
Control: placebo ($n=220$); treatment: prednisone ($n=226$).
Follow-up times are in years.  HC settings: $K=50$ intervals,
$n_\mathrm{perms}=2000$.

\begin{table}[h]
  \centering
  \caption{CSL liver-cirrhosis OS: all competing methods.}
  \label{tab:csl}
  \begin{tabular}{@{}lrr@{}}
    \toprule
    Method & Statistic & $p$-value \\
    \midrule
    Log-rank                         &  0.757 & 0.384 \\
    Gehan-Wilcoxon                   &  0.439 & 0.507 \\
    Tarone-Ware                      &  0.774 & 0.379 \\
    Peto-Prentice                    &  0.574 & 0.449 \\
    Fleming-Harrington (1,1)         &  2.234 & 0.135 \\
    MaxCombo                         &  1.495 & 0.236 \\
    Yang--Prentice                   &  0.872 & 0.393 \\
    HCHG                             &  1.257 & 0.002\sig \\
    \bottomrule
  \end{tabular}
\end{table}

\section{Cross-dataset summary}

Table~\ref{tab:summary} condenses the log-rank, MaxCombo, Yang--Prentice,
and HCHG $p$-values reported in the main text.  Full statistics for all
eight methods appear in Section~4.

\begin{table}[h]
  \centering
  \caption{Summary of log-rank, MaxCombo, Yang--Prentice, and HCHG
           $p$-values across the four case studies (full tables in Section~4).}
  \label{tab:summary}
  \small
  \begin{tabular}{@{}llrrrrr@{}}
    \toprule
    Dataset & Endpoint & $n$ & Log-rank & MaxCombo & Yang--Prentice & HCHG \\
    \midrule
    CheckMate~057  & PFS &  582 & 0.351 & $2.5\times 10^{-5}$ & 0.00055 & 0.001 \\
    COMET-1        & OS  & 1028 & 0.262 & 0.338 & 0.262 & 0.014 \\
    AZURE          & DFS & 3359 & 0.305 & 0.401 & 0.308 & 0.005 \\
    CSL            & OS  &  446 & 0.384 & 0.236 & 0.393 & 0.002 \\
    \bottomrule
  \end{tabular}
\end{table}

\noindent
Across all four datasets, HCHG achieves $p \leq 0.014$ while the log-rank
test is non-significant throughout ($p \geq 0.26$). MaxCombo and
Yang--Prentice detect the delayed-benefit crossing on CheckMate~057 but
are non-significant on COMET-1, AZURE, and CSL ($p \geq 0.24$). Among the
single weighted log-rank alternatives, only CheckMate~057 Gehan-Wilcoxon
and Peto-Prentice reach $p < 0.05$ (at 0.041 and 0.049 respectively).

\section{Machine-readable results}

The CSV files below contain the same values as Tables~\ref{tab:checkmate}--\ref{tab:csl}
(columns \texttt{method}, \texttt{statistic}, \texttt{p\_value}):

\begin{itemize}
  \item \texttt{AppNote/results/immuno\_test\_results.csv}
  \item \texttt{AppNote/results/comet\_test\_results.csv}
  \item \texttt{AppNote/results/azure\_test\_results.csv}
  \item \texttt{AppNote/results/csl\_test\_results.csv}
  \item \texttt{AppNote/results/maxcombo\_yp\_details.json}
        (MaxCombo component $Z$-statistics and Yang--Prentice
        $\hat\theta_1$, $\hat\theta_2$).
\end{itemize}

\bibliographystyle{plainnat}
\bibliography{refs}

\end{document}